\documentclass[aps,pre]{revtex4}

\usepackage{graphicx}
\usepackage{color}

\def\be{\begin{equation}}
\def\ee{\end{equation}}     
\def\bfi{\begin{figure}}
\def\efi{\end{figure}}
\def\bea{\begin{eqnarray}}
\def\eea{\end{eqnarray}}

\begin{document}

\title{Scaling in the Aging Dynamics of the Site-diluted Ising Model}

\author{Federico Corberi}
\affiliation {Dipartimento di Fisica ``E.~R. Caianiello'', and INFN, Gruppo Collegato di Salerno, and CNISM, Unit\`a di Salerno,Universit\`a  di Salerno, 
via Ponte don Melillo, 84084 Fisciano (SA), Italy.}

\author{Eugenio Lippiello}
\affiliation{Dipartimento di Scienze Ambientali, Seconda Universit\`a di Napoli,
Via Vivaldi, Caserta, Italy}

\author{Anupam Mukherjee}
\affiliation{CIICAp, UAEM, Av. Universidad 1001 Col. Chamilpa, Cuernavaca, Morelos 62210, Mexico}

\author{Sanjay Puri}
\affiliation{School of Physical Sciences, Jawaharlal Nehru University, New Delhi 110067, India}

\author{Marco Zannetti}
\affiliation {Dipartimento di Fisica ``E.~R. Caianiello'', and INFN, Gruppo Collegato di Salerno, and CNISM, Unit\`a di Salerno,Universit\`a  di Salerno, via Giovanni Paolo II 132, 84084 Fisciano (SA), Italy.}

\begin{abstract}
We study numerically the phase-ordering kinetics of the two-dimensional site-diluted Ising model. The data can be interpreted in a framework motivated by renormalization-group concepts. Apart from the usual fixed point of the non-diluted system, there exist two disorder fixed points, characterized by logarithmic and power-law growth of the ordered domains. This structure gives rise to a rich scaling behavior, with an interesting crossover due to the competition between fixed points, and violation of superuniversality.
\end{abstract} 

\maketitle

\section{Introduction}

Relaxation through domain coarsening is a well established paradigm of slow non-equilibrium dynamics.
Typically, it takes place in the phase-ordering process following the
quench of a system, like a ferromagnet or a binary mixture, to below the critical point~\cite{bray,2dinoirf}.
The key feature of coarsening is the unbounded growth of the size $L(t)$ of ordered domains.
For large enough times, the existence of a dominant length scale produces an interesting aging-scaling
phenomenology in various observable quantities~\cite{zan,BCKM,ioeleti}.
The simplicity of the structure is very attractive and is expected to be valid beyond the 
realm of disorder-free phase-separating systems. 
In recent years, this has prompted a considerable effort to understand
the role of disorder in systems where its presence 
does not prevent phase-ordering~\cite{sp04}. 
Systems of this type are disordered ferromagnets, namely systems where disorder
coexists with the low-temperature ferromagnetic order. The unifying theme in this area has been the investigation of the modifications produced by disorder on the properties of the underlying pure systems
\cite{sp04,pcp91,bh91,puripowerlaw,hp06,sab08}. 
 
In our previous works, devoted to the {\it random-bond Ising model} (RBIM) and the {\it random-field Ising model} (RFIM) \cite{decandia,EPL,CLMPZ,variousnoi2},
we have addressed two long-debated issues. The first one was about the nature of the
asymptotic growth law: Power law, with a disorder dependent exponent, against logarithmic behavior.
The second was about the so-called {\it superuniversality hypothesis}, according to which
disorder affects the growth law, but not
the scaling functions of quantities like correlation and response functions~\cite{10noirf}.
We have been able to clarify both issues, presenting evidence for an asymptotic logarithmic
growth law and for the absence of superuniversal behavior. 
The achievement of these results has
required, in addition to a considerable
numerical effort, extensive use of  
a scaling approach based on the competition between pure and disorder-controlled behavior.
The overall picture emerging from our work suggests 
the existence of an underlying framework, independent of the source of disorder,
whose gross common features are: (i) disorder slows down coarsening, producing logarithmic asymptotic behavior; and (ii) the {\it relevance} of disorder, in a sense to be made precise below, excludes the validity of the superuniversality hypothesis. Differences from system to system arise only in the quantitative details.

In this context, we undertook a study of the two-dimensional {\it site-diluted Ising model} (SDIM), expecting further confirmation of the above pattern. However, our simulations have revealed a richer phenomenology which is {\it qualitatively} different from that observed in the RBIM and the RFIM. The major novelty is that, in the SDIM case, more disorder produces slower growth only when disorder is sufficiently small. If the amount of disorder goes beyond a certain threshold, more disorder produces {\it faster} growth. This lack of monotonicity does not fit into the crossover pattern as observed in the RBIM and RFIM. However, the pieces of what otherwise would be an intricate puzzle, fall into place if the scaling framework is adequately generalized. Scaling has turned out to be
an indispensable tool, without which sorting out the SDIM phenomenology would have been a difficult task.

This paper is organized as follows. The model and the simulations are presented in Section~\ref{SDIM}. The study of the growth law and its scaling analysis are carried out in Sections~\ref{GL} and~\ref{scal}. Section~\ref{autocorr} is devoted to the scaling analysis of the autocorrelation function. Finally, in Section~\ref{concl} the conclusions are presented.

\section{Site-diluted Ising Model}
\label{SDIM}

\subsection{Substrate}
\label{sub}

The substrate is prepared by generating configurations of occupied sites as in random
percolation. On the sites of a square two-dimensional lattice there are independent
random variables $n_i$, which take values $n_i=1$ (occupied site) with probability $p$
and $n_i=0$ (empty site) with probability $d=1-p$. The substrate is formed by the set of
occupied sites. In the following, $d$ will be referred to as dilution.

Let us follow the evolution of the geometrical structure of the substrate as dilution is varied from high to low values. We denote the percolation threshold by $p_c$. Then, for $d > d_c = 1-p_c \simeq 0.4072$, the substrate is formed by finite clusters of occupied sites. Their characteristic size is
\be
\xi(d) \sim |d_c-d|^{-\nu} ,
\label{sub.2}
\ee
where $\nu=4/3$ and the fractal dimension is $d_f=91/48$ \cite{Stauffer}. The dilution range $d>d_c$ will not be considered in the following since, in the absence of an infinite cluster, coarsening cannot be sustained. At $d_c$, $\xi(d)$ diverges and an infinite {\it spanning cluster} is formed. As $d$ is lowered below $d_c$, the infinite cluster is fractal over distances up to $\xi(d)$, while it becomes compact over larger distances. In addition, finite clusters are also present. This is the structure of the substrate in the dilution regime $d_c \geq d \geq d^*$, where $d^*$ is the special dilution value
defined by
\be
\xi(d^*) = a ,
\label{sub.3}
\ee
and $a$ is a characteristic microscopic length, like the lattice spacing. For dilutions below $d^*$, the infinite cluster is compact over all length scales and there are no finite clusters. There remain vacancies inside the infinite cluster which are essentially single-site vacancies. Hence, in the dilution regime $d^* \geq d \geq 0$,
the average distance between pairs of vacancies defines a new characteristic length 
\be
\lambda(d) = a(d/d^*)^{-1}. 
\label{sub.4}
\ee

\subsection{Spin System}

The SDIM is obtained by putting Ising spins $\sigma _i=\pm 1$ on the substrate or, equivalently, on the whole lattice and taking an interaction Hamiltonian of the form 
\be
H=-J\sum _{\langle ij\rangle}n_in_j\sigma _i \sigma _j .
\label{ham}
\ee
Here, $\langle ij \rangle$ denotes nearest-neighbour pairs, and $J>0$ is a ferromagnetic coupling constant. The variables $\{n_i\}$ enter the problem as quenched disorder.

\subsubsection{Equilibrium States}

\begin{figure}[t]

	\vspace{1cm}

    \centering
   \rotatebox{0}{\resizebox{.5\textwidth}{!}{\includegraphics{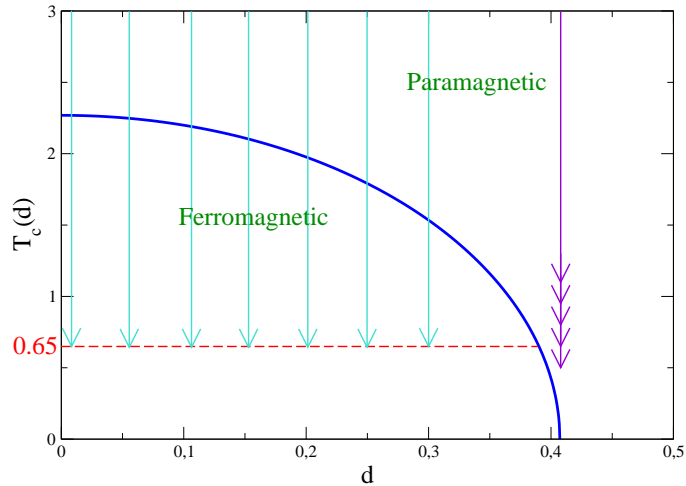}}}
   \caption{Pictorial representation of the equilibrium phase diagram of the 
SDIM. The bold blue line is the critical temperature $T_c(d)$.
   The temperature $T=0.65$ where most simulations will be performed
   is marked with a red line. Vertical arrows represent the quenching processes
   considered in the simulations.}
\label{fig_Tc}
\end{figure}

For $d \leq d_c$, at low enough temperature the system exhibits ferromagnetic order. In the $(d,T)$ plane there is a critical line $T_c(d)$, which separates the paramagnetic from the ferromagnetic phase. The equilibrium phase diagram is pictorially represented in Fig.~\ref{fig_Tc}. The critical temperature, which in the following will be measured in units of $k_B/J$, where $k_B$ is the Boltzmann constant, decreases from the pure Ising model value [$T_c(0) \simeq 2.269$] as the dilution is increased and vanishes at $d_c$ [$T_c(d_c) = 0$] since the structure is disconnected for $d>d_c$. 

\subsubsection{Time Evolution}
\label{timev}

The occupied sites evolve with non-conserved dynamics \cite{bray,2dinoirf}. We use single-spin-flip transition rates of the Glauber form
\be
w(\sigma _i \to -\sigma _i)=\frac{1}{2} \left [1-\sigma _i \tanh (H_i^W/T)\right ] .
\label{transr}
\ee
Here, $H_i^W$ is the local Weiss field obtained by the sum
\be
H_i^W = \sum _{j \in L_i} n_j \sigma_j
\ee
over the set of nearest-neighbors $L_i$ of $i$. We consider a cooling procedure where the system is prepared initially in the infinite-temperature disordered state. At the time $t=0$, it is suddenly quenched to a finite temperature $T$. We will study the following two classes of quenches: \\
(i) The dilution of the system is in the range $d \in [0,0.3]$ and the final temperature is $T=0.65$ (represented by the cyan vertical arrows in Fig.~\ref{fig_Tc}). \\
(ii) We have $d=d_c$ and different values of $T>0$ (represented by the violet arrows in Fig.~\ref{fig_Tc}).

In the first case, the system is always quenched well below the critical temperature $T_c(d)$ \cite{Ballesteros}, as sketched in Fig.~\ref{fig_Tc}. In this situation, one can speed up the simulations by using a modified dynamics where spin flips in the bulk of domains, namely those aligned with all the nearest neighbors, are prevented. This modified dynamics does not alter the behavior of the quantities we are interested in, as has been tested in a large number of cases \cite{nobulk}. We have checked that this is also true in the present study.

In the second case, the system is quenched to finite temperatures, because at $T=0$ the evolution gets frozen in
metastable states and there is no long-time dynamics. Hence we are quenching above the critical temperature $T_c(d_c)=0$ and the system will eventually relax to a disordered state with a finite spin coherence length $\xi _\sigma(T)$ [not to be confused with the substrate property $\xi (d)$]. However, since $\xi _\sigma(T)$ diverges very fast for $T\to 0$, a coarsening phenomenon is seen in the pre-asymptotic stage. The corresponding growth of $L(t)$ will eventually end at $t=t_{eq}(T)$ when $L(t)\simeq \xi _\sigma(T)$. The situation is similar to the one
found in other ferromagnetic systems with $T_c=0$ as, for instance, the one-dimensional Ising model with conserved dynamics \cite{Claudio}. In order to study the ordering phenomenon, we have set $T$ sufficiently low as to never observe the equilibration of the system. Further, it can be shown that flips of bulk spins are only observed for $t\gtrsim t_{eq}(T)$, so we use the modified dynamics discussed above in this case also.

\subsubsection{Simulation Details}

The details of the simulations are as follows. We have considered a two-dimensional square lattice system of $N=2000^2$ lattice sites. We have checked that, with this choice, no finite-size effects can be detected in the time-regime accessed by the simulations. For every choice of the parameters, we perform a certain number (in the range 10-100) of independent runs with different initial conditions and thermal histories in order to populate the non-equilibrium ensemble needed to extract average quantities that will be introduced below. 

\subsection{Observables}

The two observable quantities of interest in this paper are the typical domain size $L(t)$ and the autocorrelation function:
\be 
C(t,t_w) = \frac{1}{N} \sum_{i=1}^N \left[\left\langle \sigma_i(t) n_i \sigma_i(t_w) n_i \right\rangle
-\left\langle \sigma_i(t) n_i \right\rangle \left\langle \sigma_i(t_w) n_i \right\rangle \right] ,
\label{obs.1}
\ee
where $t, t_w$ ($0 < t_w < t$) are a pair of times after the quench. The angular brackets denote a non-equilibrium ensemble average, taken over random initial conditions and over dynamical trajectories.

For a model defined on a disconnected substrate, as it happens for $d>d^*$, phase 
ordering occurs independently on the various parts of the system and, 
correspondingly, different definitions of the growing length can be given. 
Indeed, while on the spanning cluster growth continues indefinitely, on 
finite clusters it saturates to their size. Since we are interested in the aging 
phenomenon related to the existence of a divergent length, we define 
$L(t,d)$ as the characteristic length of the ordered regions which are effectively 
growing. As we will discuss soon, this quantity can be computed
as the inverse excess energy:
\be
L(t,d)=[E(t,d)-E_\infty(d)]^{-1} .
\label{lt}
\ee
Here, $E(t,d)=\langle H(t)\rangle$ is the energy at time $t$, and $E_\infty (d)$ is the energy of the equilibrium state at the final temperature $T$. Equation~(\ref{lt}) is often used to determine $L(t)$ in non-diluted systems
\cite{bray}. It expresses the simple fact that the excess energy of the coarsening system with respect to the equilibrated one is associated with the density of domain walls. This, in turn, is inversely proportional to the typical domain size. Besides its simplicity, in the diluted case the definition (\ref{lt}) has the further advantage that the disconnected finite parts of the substrate which are already ordered do not contribute to the computation of $L(t,d)$. Indeed a finite cluster is by definition surrounded by empty sites and hence there is no excess energy associated with it when its spins are aligned. 

\section{Growth Law}
\label{GL}

\begin{figure}[t]
    \centering
   \rotatebox{0}{\resizebox{.45\textwidth}{!}{\includegraphics{compare3.eps}}}
                                          
   \rotatebox{0}{\resizebox{.45\textwidth}{!}{\includegraphics{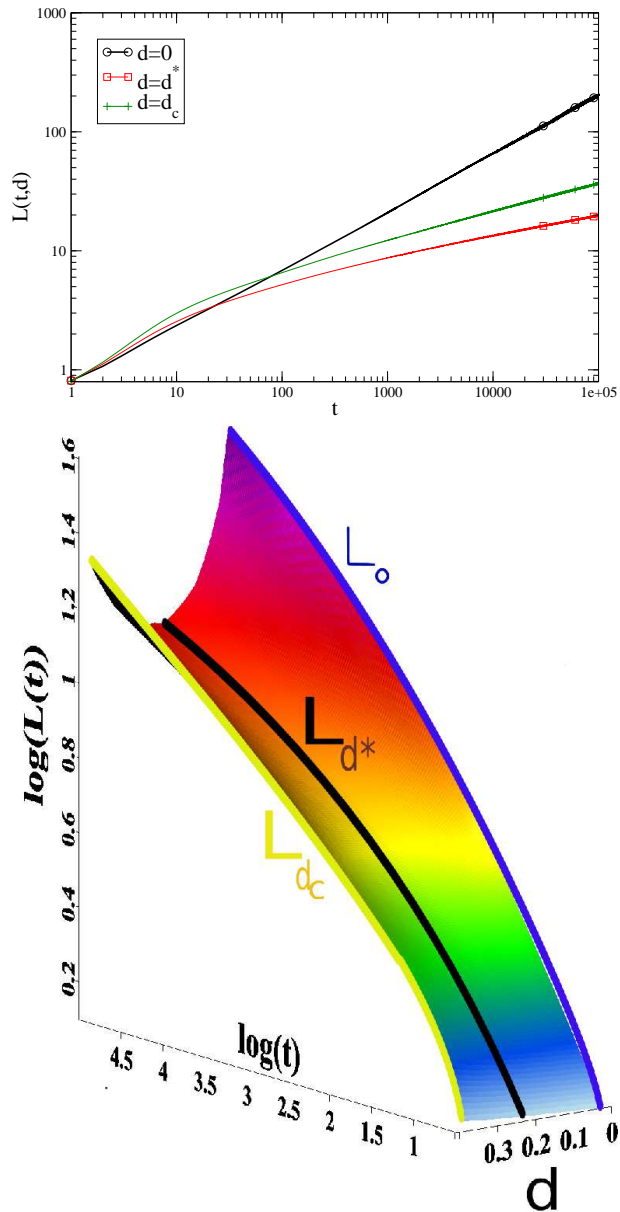}}}
   
   \caption{The behavior of the typical length $L(t)$ after a quench to $T=0.65$ is shown 
   in the upper panel for the three dilutions $d=0, d=d^*$ and $d=d_c$ in a double    logarithmic plot. In the lower panel the same quantity is shown  
   in a 
   three-dimensional plot as $t$ and $d$ are varied. Color code is from
   light blue ($L(t,d)$ very small) to violet ($L(t,d)$ very large).
   The three bold lines are the curves for 
   $d=0$, $d=d^*$, $d=d_c$ plotted in the upper panel.}
\label{fig_all}
\end{figure}

The time dependence of $L(t,d)$, for the three dilution values
$d=0,d^*=0.225,d_c$,
is plotted in the upper panel of Fig.~\ref{fig_all}. Here $d^*$ is identified
as the dilution value corresponding to the slowest numerically observed
asymptotic growth. 
The connection with the dilution $d^*$ of Eq.~(\ref{sub.3}) will be
clear below.
The plot shows that
disorder is a relevant perturbation with respect to pure behavior, 
since for $d > 0$ the late time growth is considerably
slower than the pure-like power law.
However, contrary to what is observed in the Ising model with random bonds and 
random fields \cite{pcp91,bh91,puripowerlaw,hp06,sab08,decandia,EPL,CLMPZ,variousnoi2}, Fig.~\ref{fig_all} shows, as anticipated in the Introduction, that the $d$-dependence of $L(t,d)$, at fixed time $t$, is non-monotonic. 
The nature of the problem can be grasped at glance by looking
at the lower panel of Fig.~\ref{fig_all}, which has been produced by a fine sampling
of the dilution range $[0,d_c]$. In the three-dimensional plot the growth law appears as a surface
with the shape of an upward bending valley. 
The growth-law of the pure system, denoted as $L_0(t)$ is marked
in the figure with a bold blue line. 
This is the case where
the growth is faster, as it can be clearly seen, in agreement with 
the general observation that, particularly at low $T$, 
the dynamics of a disordered system
is slower then in the corresponding  pure one, because disorder pins
the interfaces. 
Naively one could also expect that the kinetics gets slower and slower
as the parameter that controls the strength of the disorder
(in this case $d$) is increased, as it is generally observed
in various disordered ferromagnetic models 
\cite{pcp91,bh91,puripowerlaw,hp06,sab08,various2,decandia,EPL,CLMPZ,variousnoi2,pp}.
This feature is also found in the present system, but only when the dilution is increased
from zero up to $d=d^*$. The growth-law at
this density is marked by
the bold dark line at the bottom of the valley
in the figure. For $d>d^*$  growth increases again
up to the bold yellow line corresponding to $L(t,d_c)$. 
The non monotonous
character of the growth law, for fixed $t$, corresponds to descending toward the bottom 
of the valley and then climbing again as $d$ is varied.
This non-monotonicity prevents the
explanation of the data within the straightforward scaling framework arising from
the competition between an unstable pure fixed point
and a stable disorder-controlled fixed point \cite{EPL,CLMPZ,variousnoi2}.
In the next section we shall develop the scaling approach required by the above phenomenology.

\section{Scaling}
\label{scal}

From the discussion of the substrate made in subsection \ref{sub}, the three dilutions $d=0, d=d^*, d=d_c$
emerge as special values, since the asymptotic states toward which
the system evolves are scale-free, in the sense that all lengths involved are either zero
or infinite. Clearly, $L(t,d)$ diverges in all cases as $t \rightarrow \infty$. In
addition $\lambda(0) = \infty$, $\xi(d_c)=\infty$ and, at $d^*$, both
$\xi(d^*)$ and $\lambda(d^*)$ are of the order of the microscopic length $a$,
which for the sake of simplicity we shall now treat as negligible, that is $\xi(d^*) \simeq \lambda(d^*) \simeq 0$.
Therefore, the three dilutions $d=0, d=d^*, d=d_c$ are candidates 
for fixed points, in the renormalization group (RG) sense of the word, lying on the dilution axis.

If this was correct, the growth low for $d$ other than a fixed point value ought to
exhibit crossover from the pre-asymptotic behavior, characteristic of the nearby unstable
fixed point, to the asymptotic behavior characteristic of the nearby stable
fixed point. Indeed, this is what happens. 

\subsection{Small Dilution: $0 \leq d \leq d^*$}

\begin{figure}[t]
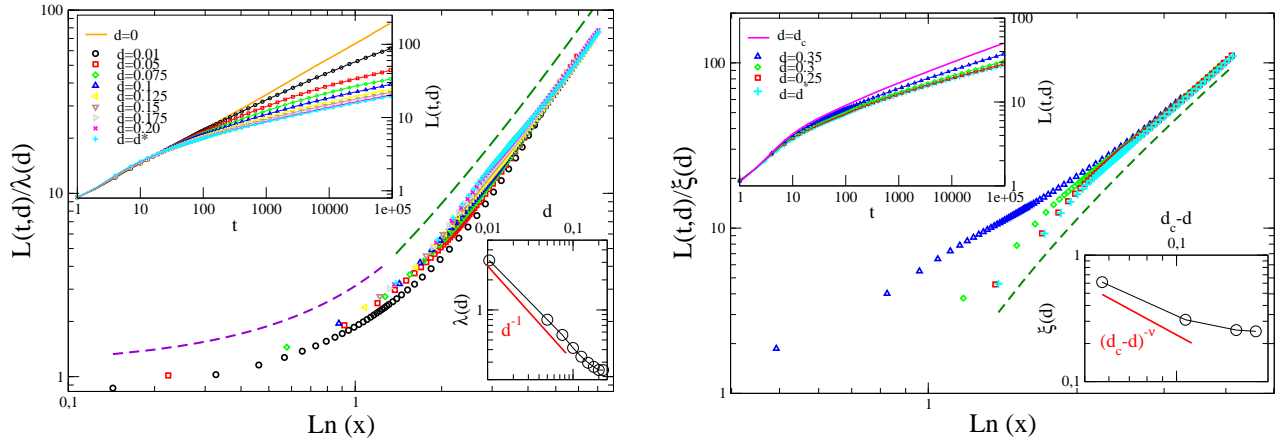

\vspace{1cm}
    \centering
   \rotatebox{0}{\resizebox{.45\textwidth}{!}{\includegraphics{fig_scaling_below_new.eps}}}
   \hspace{.5cm}
      \rotatebox{0}{\resizebox{.45\textwidth}{!}{\includegraphics{fig_scaling_above_new.eps}}}
   \caption{In the upper insets the raw data for $L(t,d)$ are plotted
   for quenches at $T=0.65$ and various $0\leq d\leq d^*$ (left panel)
   or $d^*\leq d\leq d_c$ (right panel) in a log-log plot.
   In the main figures the scaling functions $\ell $ (left panel)
   and ${\cal L}$ (right panel) are plotted in a log-log plot against $\ln x$,
   where $x=t^{1/z}/\lambda (d)$
   (left panel) or $x=t^{1/\zeta (T)}/\xi (d)$ (right panel).
   The dashed violet and green lines are the fitting curves 
   $y\propto x$ and $y=(a+b\ln x)^{1/\psi}$
   with $x=t^{1/z}/\lambda (d)$ and the values of $a,b,\psi$ given in the text
   (the curves have been vertically displaced for clarity).
   The lower insets show the dependence of the rescaling parameter
   $\lambda (d)$ and $\xi (d)$ on $d$. The bold red lines are the expected behaviors,
respectively Eq.~(\ref{sub.4}) and Eq.~(\ref{sub.2}).}
\label{fig_scaling}
\end{figure}

Let us consider first the dilution regime
$0 \leq d \leq d^*$, which will be referred to as small dilution. 
Assuming scaling and recalling that in the pure
case 
\be
L(t,0) \sim t^{1/z}
\label{scal.0}
\ee
with $z=2$, the growth law can be written in the form
\be
L(t,d) = \lambda (d)\,\ell \left (\frac{t^{1/z}}{\lambda (d)} \right ).
\label{scal.1}
\ee
According to the assumption, the data obtained for
different values of $d$ should collapse on a single master curve $\ell(x)$ by plotting
$L(t,d)/\lambda (d)$ against $x =t^{1/z}/\lambda(d)$. This is shown in the left panel of Fig.~(\ref{fig_scaling}). 
The upper inset displays the raw data in a log-log plot: The upper straight line is the power law~(\ref{scal.0})
of the pure case. As $d$ is increased from zero, growth slows down and converges toward $L(t,d^*)$, which is the slowest growth. Extracting empirically the length $\lambda (d)$ (lower inset) the data can be collapsed, as depicted in the main panel.

The master curve in the left panel of Fig.~\ref{fig_scaling}
grows linearly in $x$ for $1\gtrsim x$ and  
can be fitted with the curve
$y=(a+b\ln x)^{1/\psi}$ with $a=0.3232$, $b=0.5325$, and $1/\psi =2.3545$
in the region $x\gtrsim 1$.
In conclusion
\be
\ell(x) \sim \left \{ \begin{array}{l}
(\ln x)^{1/\psi}
\hspace{.5cm},\hspace{.5cm}\mbox{for } x \gg 1,  \\
x
\hspace{.5cm},\hspace{.5cm}\mbox{for } x \ll 1. \\
\end{array}
\right .
\label{scal.2}
\ee
This shows that the pure-like behavior of Eq.~(\ref{scal.0}) is unstable to disorder perturbation
and that the $d^*$-behavior 
\be
L(t,d^*)\sim \lambda (d)\, (\ln t)^{1/\psi},
\label{logb}
\ee
with $1/\psi \simeq 2.35$, is asymptotically dominant for all $d$ in the 
rangle $(0,d^*]$. The logarithmic behavior (\ref{logb}) is in contrast with the
results of Ref.~\cite{puripowerlaw}, where data are interpreted as power law growths
for any $d$, while it agrees with results for small dilutions of Ref.~\cite{pp}
(although a slightly larger exponent $1/\psi \simeq 2.5$ is reported). 
The discussion above shows that $d=0$ and $d^*$ can be identified, respectively, with the unstable 
and the stable fixed point in the small dilution regime. 
Notice (lower inset of the left panel in Fig.~\ref{fig_scaling}) that
$\lambda (d) \sim d^{-1}$ as $d \rightarrow 0$, in agreement with
the behavior expected from Eq.~(\ref{sub.4}), and that the the collapse is not perfect. The spread of
the curves for $x \ll 1$ reveals the existence of a correction to scaling due to the microscopic length $a$ introduced in Eq.~(\ref{sub.3}). $a$ enters 
the scaling function of Eq.~(\ref{scal.1}) through an extra variable 
$t^{1/z}/a$, which does not appears explicitly in order to keep a compact notation. 
Since generally $a\ll \lambda (d)$ this term produces corrections
at small times, as we observe in Fig.~\ref{fig_scaling}.

\subsection{Large Dilution: $d^* \leq d \leq d_c$}

\begin{figure}[t]
	\vspace{1cm}
    \centering
   \rotatebox{0}{\resizebox{.95\textwidth}{!}{\includegraphics{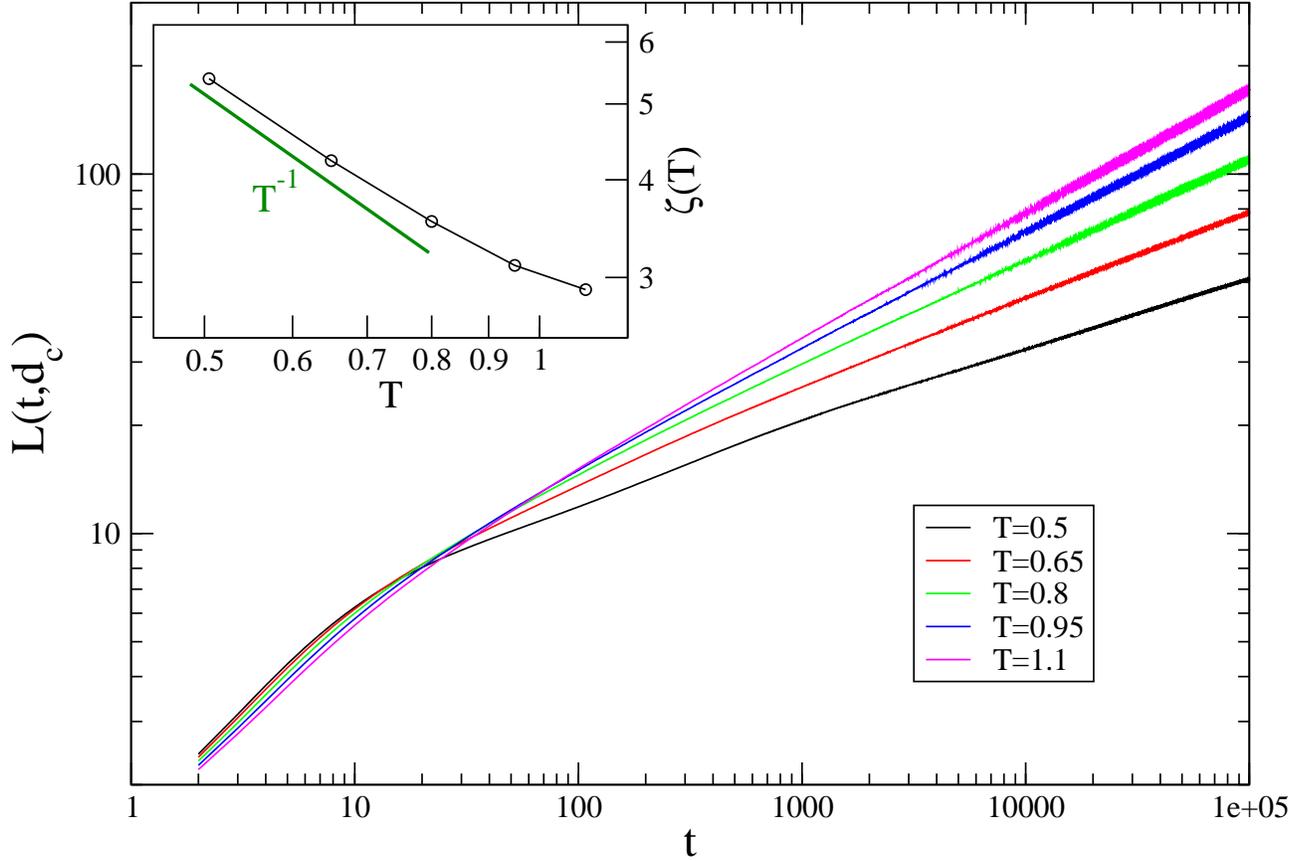}}}
   \caption{$L(t,d_c)$ is plotted against $t$ in a quench 
at different temperatures (see key).
In the inset $\zeta (T)$ (obtained as the best fit of the curves
in the large-$t$ sector) is plotted against $T$ in a double logarithmic
graph. The bold green line is the law $T^{-1}$.} 
\label{growth_law_dc}
\end{figure}

Let us first of all recall the discussion of Sec. \ref{timev} according to which
all the quenches with $d<d_c$ are below the critical temperature while in the quench with $d=d_c$ we consider the pre-asymptotic phase-ordering stage
preceding equilibration in a disordered state.
This being said, a scaling analysis analogous to the previous one can be carried out in the large dilution regime $d^* \leq d \leq d_c$. 
Looking at the raw data in the upper inset of the right panel of Fig.~\ref{fig_scaling},
the fastest growth occurs at $d_c$, slowing down
as $d$ decreases and
converging toward $L(t,d^*)$, which is again the slowest growth. Therefore, in  the large dilution regime the
unstable fixed point is at
$d_c$, while $d^*$ is still the attractive fixed point, being stable with respect to perturbations from
both the small and the high dilution side.
As shown in Fig.~\ref{growth_law_dc}, right at $d_c$ the growth law
obeys the power law (see Fig.~\ref{growth_law_dc})
\be
L(t,d_c) \sim t^{1/\zeta(T)} ,
\label{pl.1}
\ee
where $\zeta(T) > 2$ is a temperature-dependent exponent. For low temperatures, this behaves as 
\be
\zeta (T) \sim T^{-1}.
\label{expzeta}
\ee
An argument to explain
these facts will be presented below in Sec.~\ref{anarg}.

Recalling that in the large dilution regime the characteristic length is given by $\xi(d)$,
the scaling form of the growth law, analogous to Eq.~(\ref{scal.1}), is given by
\be
L(t,d) = \xi (d){\cal L}\left (\frac{t^{1/\zeta(T)}}{\xi (d)} \right ) ,
\label{pl.2}
\ee
where the chosen value of the final temperature $T$ enters explicitly through $\zeta(T)$.
This is checked following the same procedure as above. After extracting the length 
$\xi (d)$ 
(see lower inset in the right panel of Fig.~\ref{fig_scaling}), the collapse of the data is displayed in
the main panel, with the master curve obeying limiting behaviors analogous to those in Eq.~(\ref{scal.2})
\be
{\cal L}(x) \sim \left \{ \begin{array}{l}
(\ln x)^{1/\psi}
\hspace{.5cm},\hspace{.5cm}\mbox{for } x \gg 1,  \\
x
\hspace{.5cm},\hspace{.5cm}\mbox{for } x \ll 1, \\
\end{array}
\right .
\label{pl.3}
\ee
where now $x =t^{1/\zeta(T)}/\xi(d)$, and $1/\psi \simeq 2.35$.

\subsection{An Argument for $L(t,d)$}
\label{anarg}

\begin{figure}[t]
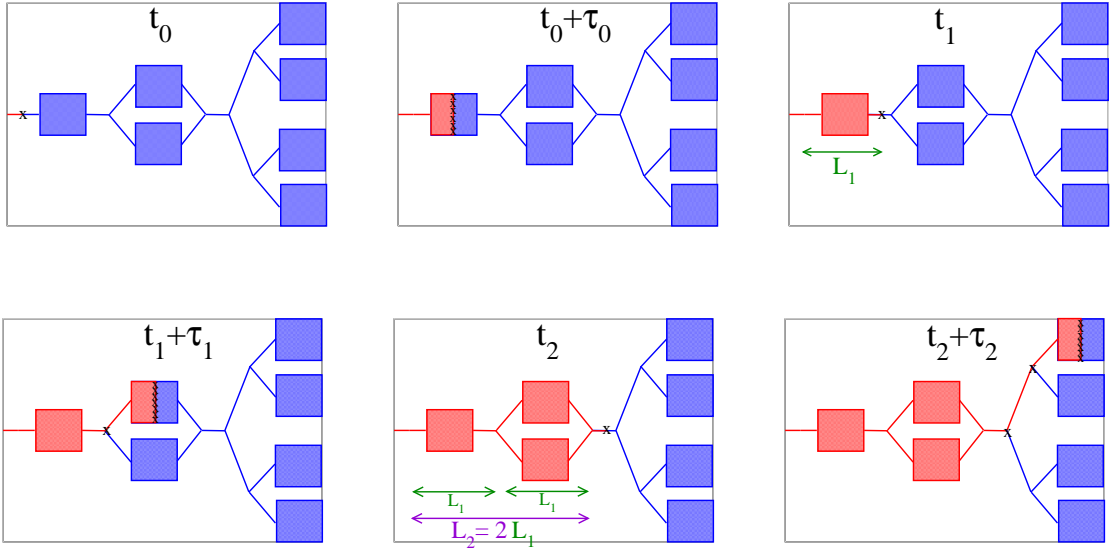

\vspace{1cm}
    \centering
   \rotatebox{0}{\resizebox{.25\textwidth}{!}{\includegraphics{fig_schema_percolaz1.eps}}}
	\hspace{.5cm}
   \rotatebox{0}{\resizebox{.25\textwidth}{!}{\includegraphics{fig_schema_percolaz2.eps}}}
	\hspace{.5cm}
   \rotatebox{0}{\resizebox{.25\textwidth}{!}{\includegraphics{fig_schema_percolaz3.eps}}}

\vspace{1cm}
   \rotatebox{0}{\resizebox{.25\textwidth}{!}{\includegraphics{fig_schema_percolaz4.eps}}}
\hspace{.5cm}
   \rotatebox{0}{\resizebox{.25\textwidth}{!}{\includegraphics{fig_schema_percolaz5.eps}}}
\hspace{.5cm}
   \rotatebox{0}{\resizebox{.25\textwidth}{!}{\includegraphics{fig_schema_percolaz6.eps}}}
   \vspace{1cm}
   \caption{Schematic representation of coarsening on the percolation network. An x 
represents an unsatisfied bond.}
\label{fig_schema_percolaz}
\end{figure}

In this section we present an argument which explains the
growth-laws observed. We start with the case $d=d_c$,
focusing the attention on the
fractal spanning cluster, 
a portion of which is schematically sketched 
in Fig.~\ref{fig_schema_percolaz}. 

The six panel represent different snapshots of the system while it is progressively
crossed by an interface. Up spins are colored in blue while the down ones are red.
Initially, at some time $t_0$ (up-left panel), the interface is located on the very left 
of the figure, and then it moves to the right at the subsequent times 
$t_0+\tau _0<t_1<t_1+\tau _1<t_2<t_2+\tau _2$ represented in the other panels. 
The cutting bonds \cite{Stauffer}, namely those links whose removal causes
disconnection,  are represented as
lines connecting {\it bulky} regions, free of cutting bonds, represented by 
squares.  
Let us denote with $L_1$ the size of the down spin domain at time $t_1$ 
and with $L_2$ that at $t=t_2$. One has $L_2=k L_1$ ($k=2$ in the 
figure) or, in general, defining $L_m$ as the size at time $t_m$, 
\be
L_m=kL_{m-1}.
\label{multl}
\ee
Analogously, by counting the number of 
down spins $N_1$ ($N_2$) at $t=t_1$ ($t_2$) one has $N_2=n N_1$
($n=3$ in the figure) and, in general, $N_m= n N_{m-1}$. This imply that 
the fractal dimension $d_f$ of the substrate defined by $N_m \propto L_m^{d_f}$
is $d_f=\ln n/\ln k$.  

We now want to study the energetic barriers that the interface has to overcome 
while moving.
The highest energy $E_1^{max}$ ($E_2^{max}$) of the interface in the 
time interval $[t_0,t_1]$ ($[t_1,t_2]$)
will occur at some intermediate time $t_0+\tau _0$ ($t_1+\tau _1$), as sketched in 
the figure. Notice that, in going from $t_1$ to $t_2$, the two bulky regions can be crossed one after the other. This imply $E_2^{max}\simeq E_1^{max}+qJ$ because at time $t_1+\tau _1$ there is an
energy $E_1^{max}$ associated to the piece of interface spanning the upper bulky region, as at $t=t_0+\tau _0$ (we make the assumption that bulky regions are equivalent) but an extra bond ($q$ bonds in a more generic network) with misaligned spins is present (the one connecting
the lower bulky region), with an associated energy $J$. 
Generalizing the above result to a generic step one has 
$E_{m}^{max}=E_m^{max}+qJ$.
The height of the energetic barrier in going 
from $t_m$ to $t_{m+1}$ is given by $\Delta E_{m}=E_m^{max}-E^{min}$,
where the minimum energy of the interface $E^{min}$ is taken 
at $t_m$ and at $t_{m+1}$ and it  
does not depend on $m$ because there is always the same number of  broken bonds in these states. This implies
\be
\Delta E_{m+1}=\Delta E_m+qJ.
\label{dem}
\ee
Expressing $m$ in terms of the size $L_m$  
through Eq.~(\ref{multl}), and dropping the index $m$ (i.e. posing $L_m=L$) we can write
$\Delta E(k L)=\Delta E(L)+qJ.$
Solving we find scaling of the barriers with $L$ as
\be
\Delta E(L)=a \ln L,
\ee
with $a=qJ/\ln k$. 
Finally, using the Arrhenius law
$t\sim e^{\Delta E/(k_BT)}$
for the time needed to exceed an energetic barrier $\Delta E$ we arrive at 
\be
L(t)\propto t^{1/\zeta(T)},
\label{llpl}
\ee
with 
\be
\zeta(T)=\frac{a}{k_BT},
\ee
which agree with Eqs.~(\ref{pl.1},\ref{expzeta}).

\begin{figure}[t]
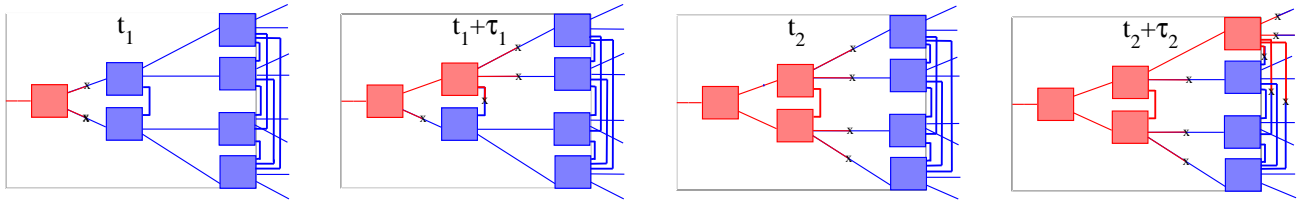

\vspace{1cm}
    \centering
   \rotatebox{0}{\resizebox{.22\textwidth}{!}{\includegraphics{fig_schema_percolaz1_bis.eps}}}
	\hspace{.3cm}
   \rotatebox{0}{\resizebox{.22\textwidth}{!}{\includegraphics{fig_schema_percolaz2_bis.eps}}}
	\hspace{.3cm}
   \rotatebox{0}{\resizebox{.22\textwidth}{!}{\includegraphics{fig_schema_percolaz3_bis.eps}}}
	\hspace{.3cm}
   \rotatebox{0}{\resizebox{.22\textwidth}{!}{\includegraphics{fig_schema_percolaz4_bis.eps}}}

   \caption{Schematic representation of coarsening on the network with $d<d_c$.}
\label{fig_schema_percolaz_bis}
\end{figure}

The situation is different for $d<d_c$, because here the cutting bonds are absent.
In order to represent schematically this property we increase the connectivity
of the substrate of Fig.~\ref{fig_schema_percolaz} in such a way that any bulky
region is directly connected by $q$ link with the others, as
shown in Fig.~\ref{fig_schema_percolaz_bis} (where $q=1$). 

Starting from a configuration of
minimum energy $E^{min}_1=2qJ$ at time $t_1$ one arrives to a maximum 
$E^{max}_1=4qJ$ at time $t=t_1+\tau _1$.
Then the new minimum with
$E^{min}_2=4qJ$ is reached at $t_2$, followed by the next maximum of energy
$E^{max}_2=8qJ$ at $t=t_2+\tau _2$. 
On a generic structure, for a generic $m$ one has $E^{min}_m=K^mqJ$
and $E^{max}_m=bK^mqJ$, where $K$ describes how the number of links
increases as the interface spans the structure 
and $b$ is a constant ($K=2$ and $b=2$ in the figure).
Hence $\Delta E_m=(b-1)K^mqJ$.
Then, due to the modified connectivity of the network, in place of
Eq.~(\ref{dem}) one has
\be
\Delta E_{m+1}=\Delta E_m +(b-1)(K-1)K^mqJ,
\ee 
leading to $\Delta E(kL)=\Delta E(L)+(b-1)(K-1)(L/L_0)^\psi qJ$, 
with $\psi = (\log _Kk)^{-1}$, 
from which
\be
\Delta E(L)=a L^\psi,
\ee
where $a=(b-1)(K-1)qJ/[L_0^\psi(K^\psi -1)]$. Using the Arrhenius law, one arrives at 
\be
L(t,d<d_c)\propto (\ln t)^{1/\psi},
\label{lllog}
\ee
in agreement with Eq.~(\ref{logb}). Notice that, in the present approach, the striking difference between the two
growth laws (\ref{llpl}) and (\ref{lllog}) is due to the different topological features of the substrate right at $d=d_c$ or for $d<d_c$, specifically due to the presence/absence of cutting bonds.

\section{Autocorrelation function}
\label{autocorr}

Scaling is expected to hold for any other observable quantity. In particular, for
the autocorrelation function~(\ref{obs.1}) we expect
\be
C(t,t_w,d)= \left \{ \begin{array}{l}
c\left ( \frac{L(t)}{L(t_w)},\frac{\lambda(d)}{L(t_w)}
\right )
\hspace{.5cm},\hspace{.5cm}\mbox{for }d\le d^*  \\
{\cal C} \left ( \frac{L(t)}{L(t_w)},\frac{\xi(d)}{L(t_w)}\right )
\hspace{.5cm},\hspace{.5cm}\mbox{for }d\ge d^*  ,
\end{array}
\right .
\label{scalc}
\ee
with the limiting forms of the scaling functions
\be
c(x,y)= \left \{
\begin{array}{l}
C_0(x) \hspace{1.5cm},\hspace{.5cm} \mbox{for } y\to \infty \\
C_{d^*}(x)\hspace{1.35cm},\hspace{.5cm} \mbox{for } y\to 0
\end{array}
\right .
\label{limc+}
\ee
and
\be
{\cal C}(x,y)= \left \{
\begin{array}{l}
C_{d_c}(x) \hspace{1.35cm},\hspace{.5cm} \mbox{for } y\to \infty \\
C_{d^*}(x) \hspace{1.35cm},\hspace{.5cm} \mbox{for } y\to 0 .
\end{array}
\right .
\label{limc-}
\ee
Here $C_0(x)$ is the thoroughly studied  
scaling function of the pure system \cite{bray, altric,paperC}, while $C_{d^*}(x)$
and $C_{d_c}(x)$ are the scaling functions at the other two
fixed points which, as far as we know, have not been studied before.
Notice the absence of powers of $L$ in front of the scaling functions.
As it is well known \cite{fracdomain}, this is due to the fact that
domains grow compactly on the substrate, even if the substrate
itself, as at the percolation threshold $d_c$, may be a fractal.

\begin{figure}[t]
\vspace{1cm}
    \centering
   \rotatebox{0}{\resizebox{.7\textwidth}{!}{\includegraphics{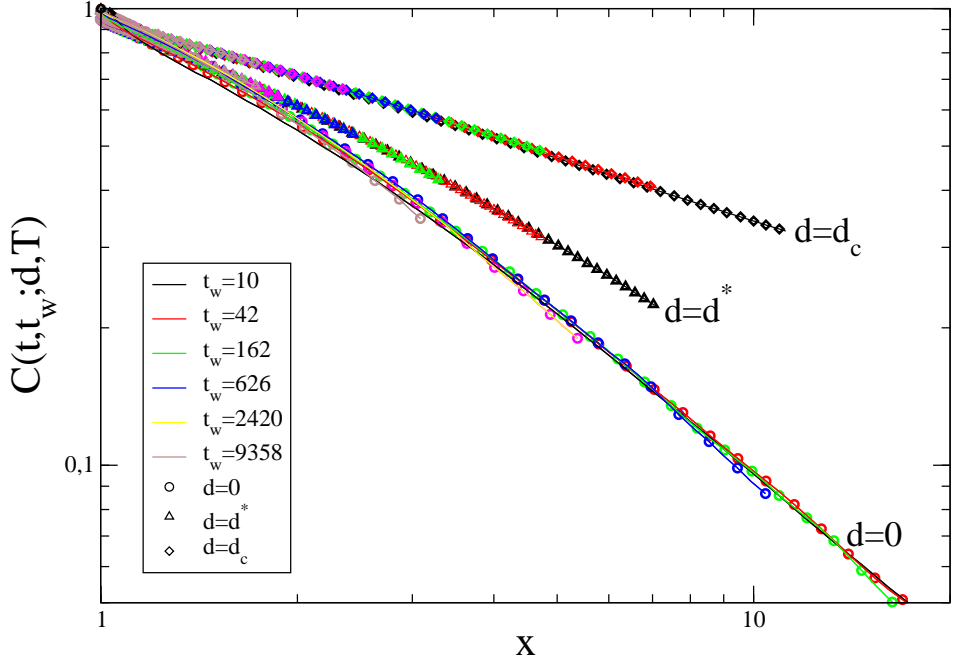}}}
\vspace{1cm}
   \caption{Data collapse of $C(t,t_w;d)$ against $x=L(t)/L(t_w)$,
for different values of $t_w$ (see caption),
for a quench to $T=0.65$ and the three fixed point dilutions
$d=0$ (lower curves), $d=d^*$ (central curves), $d=d_c$ (upper curves).}
\label{fig_scaling-c}
\end{figure}

According to the scaling forms above, if the dilution is 
set exactly at the fixed point values $d=0, d=d^*, d=d_c$,
where either $\lambda (d)=0$ or $\lambda (d)=\infty$,
or $\xi (d)=\infty$,
the autocorrelation function should depend only on $x=L(t)/L(t_w)$
when $t_w$ is
sufficiently large. This prediction is checked in  
Fig.~\ref{fig_scaling-c} where, for these three special dilutions
the expected data collapse is obtained by
plotting $C$ against $x$. In the case with $d=0$
(lower set of curves) one recovers the well known result 
\cite{bray,paperC} of  a convergence
towards data collapse in the large-$t_w$ limit. Poor 
data collapse at early $t_w$ is a feature related to
the pre-asymptotic corrections to scaling
due to the microscopic length $a$ \cite{bray,paperC}.
By contrast, at $d^*$ (set of curves in the middle) and 
at $d=d_c$ (upper set of curves), an excellent data collapses is obtained even for 
moderate values of $t_w$.
For any value of $x>1$ the fixed point scaling functions satisfy the inequality
\be
C_0(x)<C_{d^*}(x)<C_{d_c}(x).
\label{ineq}
\ee

\begin{figure}[t]
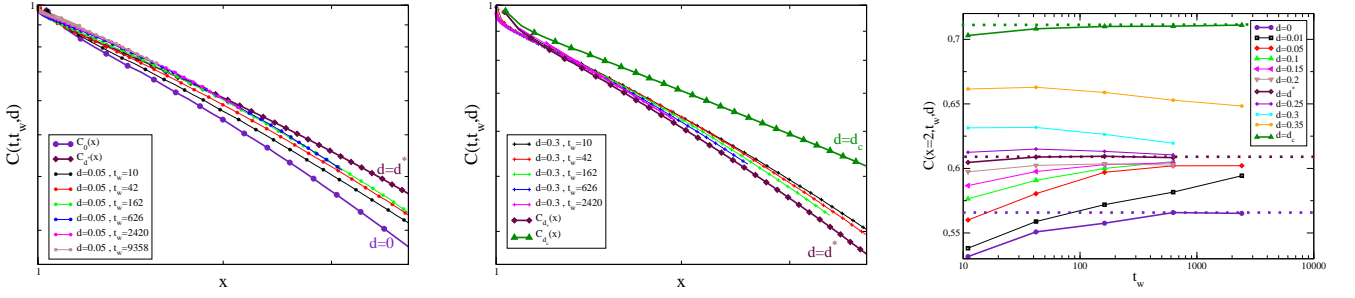

\vspace{1cm}
    \centering
   \rotatebox{0}{\resizebox{.3\textwidth}{!}{\includegraphics{fig_noscaling-c1.eps}}}
\hspace{.5cm}   
\rotatebox{0}{\resizebox{.3\textwidth}{!}{\includegraphics{fig_noscaling-c2.eps}}}
\hspace{.5cm}   
\rotatebox{0}{\resizebox{.3\textwidth}{!}{\includegraphics{fig_noscaling-c3.eps}}}
\vspace{1cm}
   \caption{$C(t,t_w;d)$ is plotted against against $x=t/t_w$
for different choices of $t_w$ (see caption) in a quench to $T=0.65$ and 
$d=0.05$ (left panel), and $d=0.3$ (central panel). 
The bold curves with heavy symbols are the scaling functions $C_{0}(x)$,
$C_{d^*}(x)$,$C_{d_c}(x)$ (see caption).
In the right panel we plot $C(t,t_w;d,T)$ against $t_w$ by fixing $x=t/t_w=2$
in quenches to different values of $d$ (see caption) spanning the entire
range $[0-d_c]$. The bold dashed lines are guide to the eye representing the
asymptotic values $\lim _{t_w\to \infty} C(x=2,t_w,d=0)$, $\lim _{t_w\to \infty} C(x=2,t_w,d=d^*)$,
$\lim _{t_w\to \infty} C(x=2,t_w,d=d_c)$ (from bottom to top).}
\label{fig_nonscaling-c}
\end{figure}

Next, let us look at dilutions other than
the fixed point values. The two cases,
the first with $0<d<d^*$ and the second with $d^*<d<d_c$, are 
depicted in Fig.~\ref{fig_nonscaling-c} (left and central panel).
Here we find that, at variance with the fixed point cases, 
the curves for different $t_w$ do not collapse. However, in both cases as 
$t_w$ is increased the curves tend to the scaling function
$C_{d^*}(x)$ of the attractive fixed point at $d^*$
(from below and from above, respectively). Given the inequality
(\ref{ineq}), these behaviors imply that $c(x,y)$ and ${\cal C}(x,y)$ 
are slowly crossing over
from the pre-asymptotic forms $C_0(x)$ and $C_{d_c}(x)$ (respectively) at
early $t_w$, to the asymptotic one $C_{d^*}(x)$ as $y$ is varied. This is a clear-cut confirmation of the 
crossover pattern uncovered from the study of the growth law
in Sec. \ref{scal}.

An alternative representation of crossover is given
in the right panel of Fig.~\ref{fig_nonscaling-c}. Here 
$C(t,t_w,d)$ is plotted against $t_w$ for fixed $x=t/t_w$
and for different dilution values spanning the
whole range $[0,d_c]$. At the three fixed point  $(d=0,d^*,d_c)$
the curves converge toward the respective asymptotic values.
For values of $d$ different from these, 
the curves are asymptotically attracted toward the
one corresponding to $d^*$. Notice that,
for values of $d$ sufficiently close to $d_c$ (as for the case $d=0.35$),
the pre-asymptotic behavior corresponding to the nearby
unstable fixed point is observed. 
Indeed, the curve initially 
initially increases (towards the plateau value of the case $d=d_c$),
and then decreases towards the value of the $d=d^*$ case.

Let us stress that, due to the presence of
the variable $y=\lambda (d)/L(t_w)$ 
(or $\xi (d)/L(t_w)$) in Eq.~(\ref{scalc})
the autocorrelation function is a function of $x=L(t)/L(t_w)$ 
only if $d$ is set to one of the fixed point densities where 
$y$ vanishes or diverges. Conversely, in
Ref. \cite{pp} collapse of the curves for the autocorrelation
is found when plotted against $x$ at any value of $d$. 
This can be perhaps understood as due to the large values of 
$t_w$ used in Ref. \cite{pp}. It must be recalled in fact that in the
large-$t_w$ limit the curves approach the master curve 
$C_{d^*}$ of  the attractive fixed point, and it may
be numerically hard to detect the dependence on $y$.
However the dependence on $y$ implied by the scaling picture
is present, as it is clearly visible in Fig.~\ref{fig_nonscaling-c}.
Moreover at least in one case the results of \cite{pp}
are obtained for a value ($d=0.2$) of the dilution so close to $d^*$
to basically probe the scaling at $d^*$ where indeed there is no
further dependence on $y$. 

Finally, notice that the scaling behavior of the autocorrelation discussed 
insofar
excludes superuniversality, since the three scaling functions
$C_0$, $C_{d^*}$ and $C_{d_c}$ are different, depend on the
disorder strength $d$, and obey the inequality (\ref{ineq}).

\section{Discussion and Conclusions}
\label{concl}

In this paper, we have studied the phase-ordering kinetics of the two-dimensional diluted Ising model. Numerical data can be consistently interpreted in a RG-inspired scaling scheme with three fixed points (FPs): an attractive FP at $d=d^*$ and two repulsive FPs at the limits $d=0$ and $d=d_c$ of the possible dilution values. This structure can be geometrically interpreted as due to the existence of two sectors separated by $d^*$: for $d<d^*$, vacancies play the role of isolated voids separated by a distance $\lambda (d) \propto d^{-1}$, whereas for $d>d^*$, the spin network has a percolative fractal structure up to distances $\xi (d)\propto (d_c-d)^{-\nu}$. At $d^*$ these two lengths become microscopic and merge, while they respectively diverge at $d=0,d_c$, providing in this way three FPs for the dynamics and an associated pattern of crossovers regulated by their attractive/repulsive character. This behavior excludes superuniversality, as has been clearly shown when discussing the properties of the autocorrelation function.

As a final observation let us comment on the fact that, at least at $d_c$, the growth law of the domains can be understood in terms of topological properties of the spin network as, in particular, the weakness of the fractal graph due to the presence of the cutting bonds. Although the argument presented in Sec.~\ref{anarg} has been developed for $L(t)$, we expect that the role played by the topology might affect other observables. This observation provides a link between the actual system and the related problem of phase-ordering on fractal structures where the importance of analogous topological properties has been pointed out \cite{parma}. One might ask if the role of topology could have important consequences also in ferromagnets with a different kind of disorder, as for instance random bonds.

\end{document}